\documentstyle[aps,prl,twocolumn,floats]{revtex}

\begin{document}
\draft
\wideabs{
\title{$d$-Wave Pairing in the Presence of Long-Range Coulomb Interactions}
\author{G\"{o}khan Esirgen,$^{1,}$\cite{email} H.-B. Sch\"{u}ttler,$^{1}$
and N.~E. Bickers$^{2}$}
\address{$^1$Center for Simulational Physics,
Department of Physics and Astronomy,
University of Georgia, Athens, Georgia 30602--2451\protect\\
$^2$Department of Physics and Astronomy, University of
Southern California, Los Angeles, California 90089--0484}
\date{\today}
\maketitle
\begin{abstract}
The one-band extended Hubbard model in two dimensions
near band-filling $1\over2$
is solved in the fluctuation exchange approximation,
including the long-range ($1/r$) part of the Coulomb interaction,
up to 4th neighbor distance.
Our results suggest that $d_{x^2-y^2}$ pairing
in the Hubbard model is robust against the inclusion of long-range
Coulomb interactions with moderate 1st neighbor repulsion
strength $V_{1}$. 
%R2
$d_{x^2-y^2}$ pairing is suppressed 
only at large $V_{1}$ ($\gtrsim 0.25 U$--$0.4 U$),
due to incipient charge density wave instabilities.
%R2
\end{abstract}
}
%
% body of paper here
%
\def\dxy {d_{x^2-y^2}}
\def\tc {T_{\rm c}}
\def\etal {{\em et~al.}}

Several cuprate superconductors appear to exhibit a pairing 
state of $\dxy$ symmetry~\cite{d_wave_exp}.
Such a non-$s$-wave state was also found early on
in microscopic pairing theories, based on antiferromagnetic (AF)
spin correlations in the two-dimensional 
(2D) Hubbard model~\cite{d_wave_theory}.
%\cite{BiScWh,GrJoRi,d_wave_theory}.
While incorporating the local on-site $U$
Coulomb repulsion, the Hubbard-based $\dxy$ pairing theories have
so far largely ignored the effects of the spatially more extended
Coulomb matrix elements.
Although unaffected by the on-site $U$, non-$s$-wave 
pairing states can, in principle, be severely suppressed by
the longer range Coulomb repulsion, due to the spatially extended
nature of their Cooper pair wavefunctions.
%R2
For any 
%proposed microscopic 
pairing theory it is therefore
of central importance to establish whether 
non-$s$-wave pairing survives the extended
%Coulomb 
repulsion.
%R2

%R2
%In the present paper 
Here, we address this question in
a self-consistent diagrammatic framework, 
the fluctuation exchange (FLEX) approximation~\cite{flex},
applied to an {\it extended} Hubbard model.
We show that $\dxy$ pairing is (i)~robust against extended 
Coulomb terms of realistic strength and
%. The $\dxy$ pairing is 
(ii)~suppressed only if the extended repulsion becomes so strong
that it induces charge density wave (CDW), rather than
AF spin density wave (SDW) instabilities in the $1\over2$-filled
Hubbard system. This CDW scenario is
apparently 
not realized in the undoped cuprates~\cite{d_wave_theory}.
%R2

We start from the 2D extended Hubbard Hamiltonian,
%%%%
%
\begin{equation}
{\cal H}=\sum_{ ij}
\Big[
-\sum_{\sigma} t_{ij}
c^{\dagger}_{i\sigma} c_{j\sigma}
+
\frac{1}{2}\sum_{\sigma\sigma'}  V_{ij}
n_{i\sigma} n_{j\sigma'}
\Big]\:,
\label{eq1}
\end{equation}
%
%%%%
where $c^{\dagger}_{i\sigma}$ creates a hole 
with spin $\sigma$ at site ${\bf R}_{i}$
on an $N\!=\!L\!\times\!L$ square lattice with
periodic boundary conditions, and
$n_{i\sigma}=c^{\dagger}_{i\sigma}c_{i\sigma}$.
%and ${\bf \Delta R}_{ij}\!=\!{\bf R}_i\!-\!{\bf R}_j$.
%
In $t_{ij}$, we include only a 1st neighbor hybridization $t$
and the chemical potential $\mu$.
The Coulomb matrix element $V_{ij}\!\equiv\!V({\bf R}_{i}-{\bf R}_{j})$ 
comprises the Hubbard on-site repulsion 
$U\equiv V({\bf 0})$ and an extended $1/r$-part,
$V({\bf \Delta R})= V_{1}/|{\bf \Delta R}|$ for 
$0<|{\bf \Delta R}| \leq r_n$,
up to the $n$th neighbor distance $r_n$. Here
$V_{1}$ denotes the strength of the 
1st neighbor repulsion and ${\bf \Delta R}$ is
%measured 
in units of the lattice constant $a$
with $a\!\equiv\!1$ in the following.
For the cuprates, we assume 
$t\!\sim\!0.3$--$0.5\;{\rm eV}$ and $U/t\!\sim\!8$--$12$~\cite{HSSJ,ScFe}.
Absorbing dielectric screening effects from the
insulating non-Hubbard electron background
into an effective $V_{1}$, one estimates
$V_{1}/t\!\lesssim\!2$--$3$~\cite{SGEH}.
Additional screening from phonons reduces this to 
$V_{1}/t\!\lesssim\!0.3$--$0.5$~\cite{SGEH},
%However, we caution that such phonon effects 
but should 
really be treated explicitly in a more realistic 
electron-phonon model,
because of the low ({\it i.e.}, potentially relevant)
phonon energy scale.

\begin{figure}[hbt]
\begin{picture}(3.375,4.500)
\includegraphics{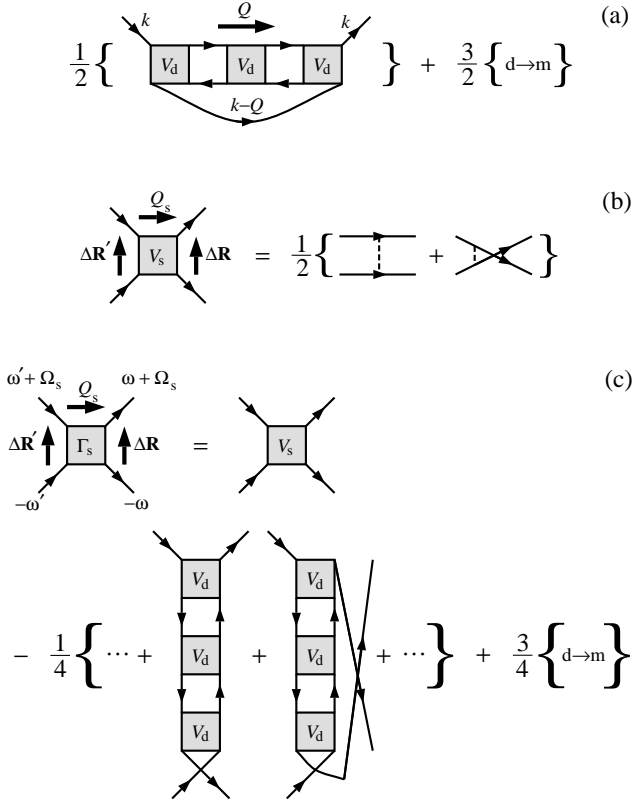}
\end{picture}
%
%%%%\begin{figure}[hbt]
%%%%\begin{picture}(3.375,4.500)
%%%%\special{psfile=fig1a.eps
%%%hoffset=-21 voffset=174 vscale=40 hscale=40 angle=0}
%%%%\special{psfile=fig1b.eps
%%%hoffset=-7 voffset=102 vscale=40 hscale=40 angle=0}
%%%%\special{psfile=fig1c.eps
%%%hoffset=-41 voffset=-20 vscale=40 hscale=40 angle=0}
%%%%%\special{psfile=fig1d.eps hoffset=-70 voffset=70
%%%%%vscale=40 hscale=40 angle=0}
%%%%\end{picture}
%
%%%%
%
\caption
{
Diagrams included in the FLEX approximation:
(a) 
Representative higher (3rd) 
order contribution to the fluctuation self-energy,
$\Sigma (k)$. 
%
%%%%%
%
(b) 
Bare singlet pairing vertex, 
$V_{\rm s}(Q_{\rm s};{\bf \Delta R};{\bf \Delta R'})$
with total pair momentum-energy 
$Q_{\rm s}\equiv({\bf Q}_{\rm s}, \Omega_{\rm s})$.
(c) 1st and typical higher (3rd) order contributions to the
full singlet pairing vertex,
$\Gamma_{\rm s}(Q_{\rm s};{\bf \Delta R},\omega;
                          {\bf \Delta R'},\omega'$).
}
\label{fig1}
\end{figure}

In Fig.~\ref{fig1}(a), we show typical (3rd order) FLEX contributions
%%%%
%
% FIG. 1:
%
%
%%%%
to the single-particle self-energy $\Sigma$, written in terms of the
bare particle-hole vertices $V_{\rm d}$ for 
density and $V_{\rm m}$ for magnetic
fluctuation exchange. In Fig.~\ref{fig1}(b) and (c), we show the bare
vertex $V_{\rm s}$ and, respectively, the 1st and typical higher (3rd) order 
contributions to the renormalized vertex $\Gamma_{\rm s}$ for the singlet 
particle-particle interaction.
We do not include particle-particle fluctuation
exchange in $\Sigma$. Hence, within our FLEX approximation~\cite{flex},
no Aslamazov-Larkin type  diagrams~\cite{flex,AsLa}
contribute to $\Gamma_{\rm s}$ at or above $T_{\rm c}$
(see also previous applications in
Ref.~\onlinecite{PaBi}).

The finite cutoff,
$V({\bf \Delta R})\equiv 0$ for 
$|{\bf \Delta R}| > r_n$,
makes a numerical solution of the FLEX equations~\cite{flex}
feasible. The key algorithmic ingredients are (i)~introduction of
a mixed real-space and momentum-space basis set~\cite{EsBi,HaSh}
for two-body propagators;
and (ii)~application of a numerical renormalization 
group to calculate Matsubara frequency sums~\cite{PaBi}.
The vertices 
($V_{\rm d}$, $V_{\rm m}$, $V_{\rm s}$)
%, $\Gamma_{\rm s}$) 
%
are $M\!\times\!M$ matrices, where $M$ 
%equals 
is the number of 
%lattice vectors 
${\bf \Delta R}$ with 
$V({\bf \Delta R})\!\neq\!0$.
For example, in the 1st neighbor ($n=1$)
Coulomb model, $M\!=\!5$ and their non-zero matrix elements at
%$V_{\rm d}$, $V_{\rm m}$ and $V_{\rm s}$
%momentum-energy transfer 
$Q\equiv({\bf Q},\Omega)$ 
[see Fig.~\ref{fig1}(a) and (b)] are
%%%%
%
\begin{eqnarray}
&&
\lefteqn{V_{\rm d}(Q;{\bf\Delta R};{\bf \Delta R'})=}
\nonumber\\
&&\left\{
\begin{array}{l@{\;}l}
U+4V_{1}(\cos Q_{x}+\cos Q_{y}),
&{\bf \Delta R}={\bf \Delta R'}= {\bf 0}\\
-V_{1},
&{\bf \Delta R}={\bf \Delta R'}=\pm \hat{x}, \pm \hat{y}
%%%%\\
%%%%0 \;, 
%%%%&\text{otherwise,}
\end{array}
\right.
%%%%\nonumber\\
\label{eq2}\\
%%%%\end{eqnarray}
%
%%%%
%%%%$V_{\rm m}$ is given by
%%%%
%
%%%%\begin{eqnarray}
&&
\lefteqn{V_{\rm m}(Q;{\bf\Delta R};{\bf \Delta R'})=}
\nonumber\\
&&\left\{ 
\begin{array}{l@{\;\quad\quad
                 \quad\quad\quad\quad\quad\quad\quad}l}
-U \;,
&{\bf \Delta R}={\bf \Delta R'}={\bf 0}\\
-V_{1} \;,
&{\bf \Delta R}={\bf \Delta R'}=\pm \hat{x}, \pm \hat{y}
%%%%\\
%%%%0 \;,
%%%%&\text{otherwise}
\end{array}
\right.
\label{eq3}\\
%%%%\end{eqnarray}
%
%%%%
%%%%and $V_{\rm s}$ in Fig.~\ref{fig1}(b) is
%%%%
%
%%%%\begin{eqnarray}
&&
\lefteqn{V_{\rm s}(Q;{\bf\Delta R};{\bf \Delta R'})=}
\nonumber\\
&&\left\{ 
\begin{array}{l@{\quad\quad\quad\quad\quad\quad}l}
U \;,
&{\bf \Delta R}={\bf \Delta R'}={\bf 0}\\
V_{1}/2 \;,
&{\bf \Delta R}=\pm {\bf \Delta R'}=\pm \hat{x}, \pm \hat{y}\;.
%%%%\\
%%%%0 \;,
%%%%&\text{otherwise.}
\end{array}
\right.
\label{eq4}
\end{eqnarray}
%
%%%%

The superconducting transition occurs when $\lambda^{\rm s}\!=\!1$
where $\lambda^{\rm s}$ is the
largest eigenvalue of the integral kernel for the
particle-particle ladder series with $\Gamma_{\rm s}$
as the interaction vertex~\cite{flex} at total pair
momentum-energy $Q_{\rm s}\!=\!0$.
The $\tc$ results reported here are for $16\!\times\!16$ lattices
and agree with $32\!\times\!32$ lattice results to within $5\%$ at 
selected parameter values we have checked.

\begin{figure}[hbt]
\begin{picture}(3.375,4.200)
\includegraphics{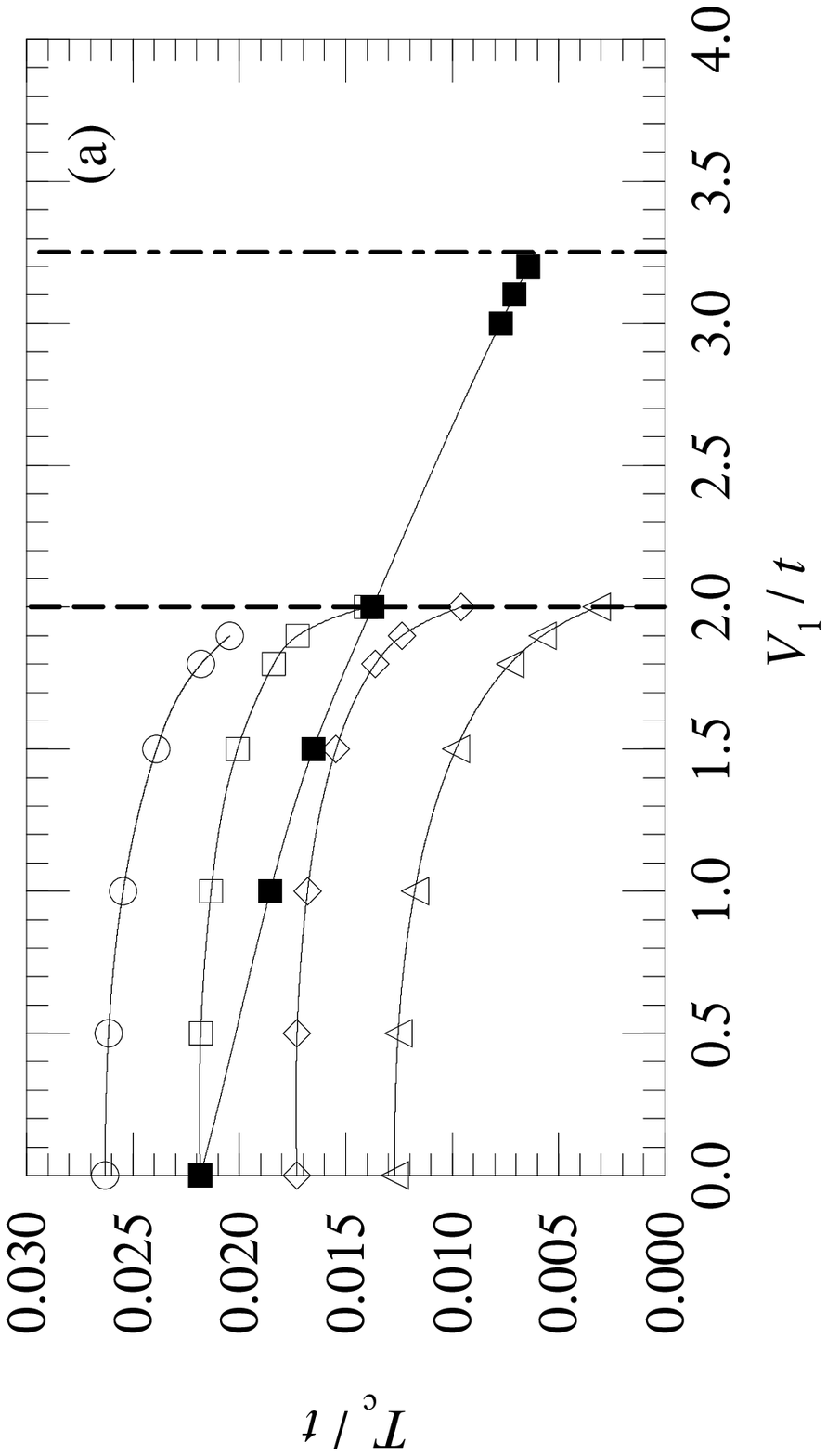}
\includegraphics{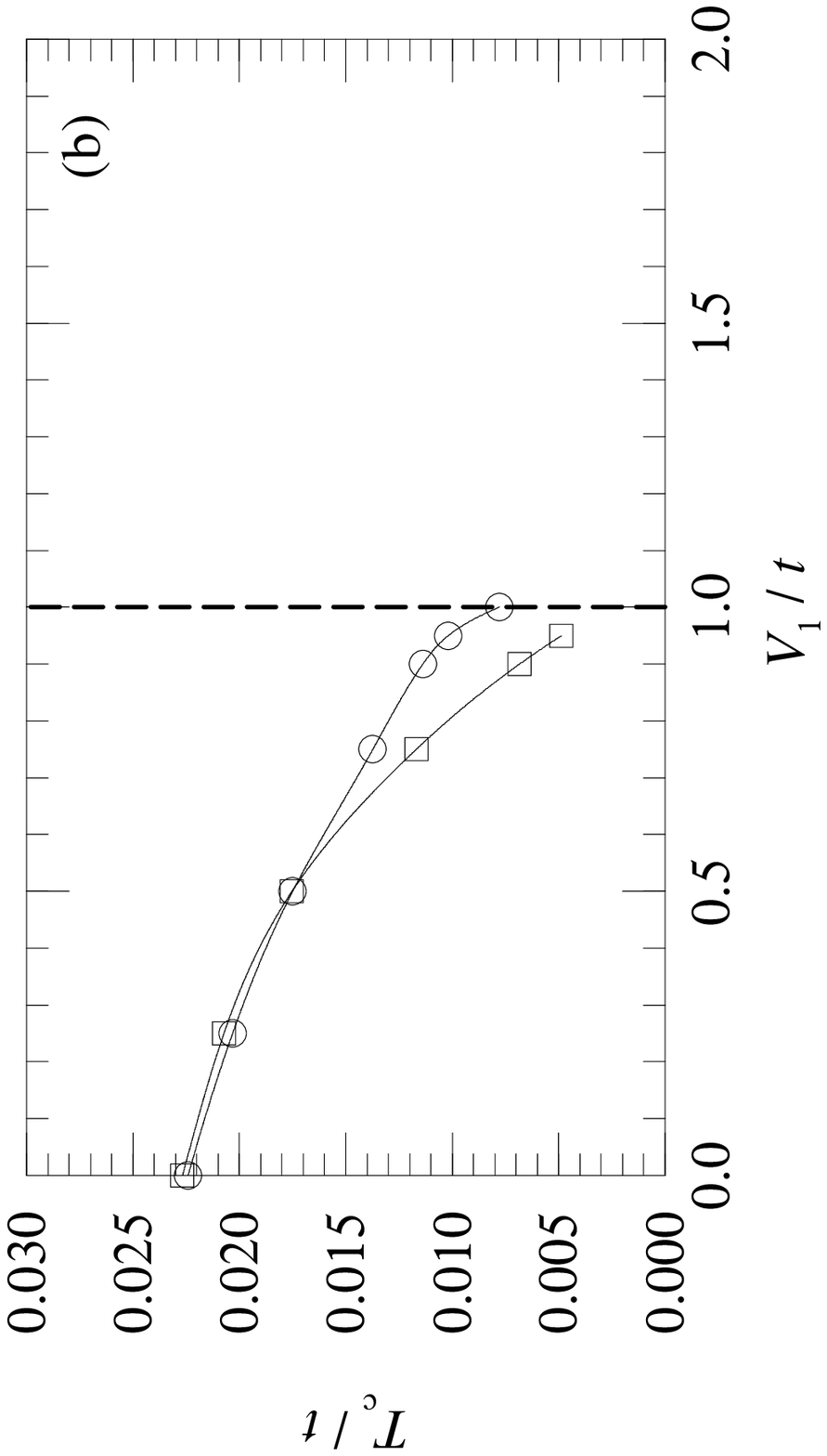}
\end{picture}
%
%%%%\begin{figure}[hbt]
%%%%\begin{picture}(3.375,5.150)
%%%%\special{psfile=fig2a.eps hoffset=-43 voffset=448 vscale=50 hscale=50
%%%%angle=-90}
%%%%\special{psfile=fig2b.eps hoffset=-47 voffset=268 vscale=50 hscale=50
%%%%angle=-90}
%%%%\end{picture}
%
%%%%
%
\caption
{
$\dxy$ transition temperature
$\tc$ vs.\ 1st neighbor Coulomb repulsion, $V_{1}$, at
(a) $U=8t$ and (b) $U=4t$ for hole densities
$\langle n \rangle=1.12$ (circles), $1.16$ (squares),
$1.20$ (diamonds), and $1.24$ (triangles).
Open symbols are for the 1st neighbor ($n=1$),
closed symbols for the 2nd neighbor ($n=2$) Coulomb model.
Vertical dashed (dot-dashed) lines mark the crossover
$V_{1}^{\rm md}$ from SDW to CDW fluctuation behavior 
in the 1st neighbor (2nd neighbor) model, as described in text.
}
\label{fig2}
\end{figure}

In Fig.~\ref{fig2} we plot the superconducting
transition temperature $\tc$ as a function 
of $V_{1}$ for different
hole concentrations and different $U$
in the 1st neighbor ($n=1$) 
and 2nd neighbor ($n=2$) 
Coulomb repulsion models.
%%%%
%
% FIG. 2:
%
%
%%%%
As in previous FLEX calculations for Hubbard-like models,
the symmetry of the dominant pairing instability is
$\dxy$~\cite{d_wave_theory,EsBi,PaBi,flex_old}.
The striking feature in Fig.~\ref{fig2} is the remarkable 
insensitivity of $\tc$ to $V_{1}$ over a wide parameter range. 
An eventual abrupt down-turn in $\tc$ occurs only when 
$V_{1}$ approaches the crossover $V_{1}^{\rm md}$
from SDW to CDW fluctuation behavior,
discussed below.

To understand this behavior, we analyze the 
decomposition of 
$\Gamma_{\rm s}$ in Fig.~\ref{fig1}(c) for the case of the
1st neighbor ($n\!=\!1$) model.
We consider
$\Gamma_{\rm s}(0;{\bf \Delta R},i \pi T;{\bf \Delta R},i \pi T)$ 
as a measure of the effective interaction
between holes separated by $\bf \Delta R$ in the $Q_{\rm s}=0$ 
singlet pairing channel at temperature $T$.
Note that $V_{1}$ affects $\tc$ 
via the 1st order (Hartree-Fock) contribution to 
$\Sigma$ (not shown in Fig.~\ref{fig1}),
via the 1st order contribution $V_{\rm s}$ 
[Fig.~\ref{fig1}(c)] to $\Gamma_{\rm s}$,
and via the charge and spin fluctuation
contributions, that is, 
via the $V_{\rm d}$- and $V_{\rm m}$-ladder 
series which enter into $\Sigma$ [Fig.~\ref{fig1}(a)]
and $\Gamma_{\rm s}$ [Fig.~\ref{fig1}(c)].

The primary reason for the robustness of the $d$-wave pairing
against moderate 1st neighbor repulsions,
$V_{1}\!\lesssim\!V_{1}^{\rm md}$, is simply
the raw strength of the effective electron-electron 
interaction generated by the spin fluctuations,
primarily with momentum transfer ${\bf Q}$ 
[see Fig.~\ref{fig1}(a)] near ${\bf Q}\!=\!{\bf (\pi,\pi)}$. 
For example, at 1st neighbor distance, $U\!=\!8t$ and
hole density 
$
\langle n\rangle\!\equiv\!
\langle \sum_{\sigma} n_{j\sigma}\rangle\!=\!1.16
$,
this attractive contribution to $\Gamma_{\rm s}$
is of order $32t$. This clearly dominates over
extended Coulomb effects at moderate $V_{1}$.

At $U\!=\!8t$, the primary effect of a moderate $V_{1}$ on
$\Gamma_{\rm s}$ is a slight enhancement of the 
1st neighbor attraction, by net amounts $\lesssim\!0.2t$.
This additional $V_{1}$-induced attraction is caused 
by ${\bf Q}\sim{\bf 0}$ charge and by
${\bf Q}\sim{\bf (\pi,\pi)}$
spin fluctuation contributions to $\Gamma_{\rm s}$,
which overscreen the direct 1st order repulsive 
$\Gamma_{\rm s}$ contribution, $V_{\rm s}$, in
Fig.~\ref{fig1}(c).
%~\cite{ov_scrn}.
The reason for the slight suppression of $\tc$ 
by a moderate $V_{1}$ at $U\!=\!8t$
is therefore {\it not} the $V_{1}$-effect on $\Gamma_{\rm s}$,
but rather its effect on $\Sigma$. 
By increasing $\Sigma$, $V_{1}$ suppresses
the single-particle spectral weight near the Fermi
energy which, in turn, tends to reduce $\tc$. This 
$\Sigma$-effect outweighs the $\Gamma_{\rm s}$-effect of $V_{1}$.

As indicated by the $U=4t$ results in Fig.~\ref{fig2}(b),
a large $U$-value is important for 
stabilizing the $\dxy$ pairing
against extended Coulomb effects.
To understand this, note first that the spin fluctuation
contributions to  $\Gamma_{\rm s}$ and $\Sigma$
are strongly $U$-dependent. For example, at $U=4t$, $V_{1}\!=\!0$ 
and $\langle n\rangle\!=\!1.16$,
the 1st neighbor attraction induced by the
spin fluctuations in $\Gamma_{\rm s}$ is reduced to $3.47t$, 
compared to $\sim\!32t$ in the $U=8t$ case.
Despite this order of magnitude decrease,
$\tc$ for $U=4t$ is slightly higher, due to the accompanying
reduction of the self-energy effects discussed above. 
However, for this $U\!=\!4t$ parameter set,
$\tc$ is also suppressed much more rapidly by
turning on $V_{1}$. This is due to the fact 
that the spin fluctuation 
contribution to the 1st neighbor attraction 
in $\Gamma_{\rm s}$ is now suppressed, rather than
enhanced, by $V_{1}$. 
As a consequence, the net effect of $V_{1}$ on $\Gamma_{\rm s}$
is to suppress the 1st neighbor attraction,
for example by about
$0.6t$ for $V_{1}\!=\!0.5V_{1}^{\rm md}$.
Due to the reduced overall strength of $\Gamma_{\rm s}$,
this is a relatively much larger effect than
in the $U=8t$ case and therefore has a much larger
effect on $\tc$.

The SDW-CDW crossover $V_{1}^{\rm md}$
in Fig.~\ref{fig2} is operationally
defined as that $V_{1}$-value where 
the maximal ``density'' eigenvalue 
$\lambda^{\rm d}$
becomes equal to the maximal ``magnetic'' eigenvalue 
$\lambda^{\rm m}$. 
Here, $\lambda^{\rm d}$ and $\lambda^{\rm m}$ denote
the largest eigenvalues of the 
integral kernels for the $V_{\rm d}$- and
%, respectively,
$V_{\rm m}$-based
particle-hole ladder series entering into
$\Sigma$ and $\Gamma_{\rm s}$, shown in
Fig.~\ref{fig1}(a) and (c).
%R2
This criterion for $V_{1}^{\rm md}$
may give only rough estimates of the
actual physical crossover since,
instead of the full interaction vertices
$\Gamma_{\rm d}$ and $\Gamma_{\rm m}$~\cite{flex},
it is obtained from the bare interaction vertices $V_{\rm d}$ and $V_{\rm m}$.
%it is not obtained from the full FLEX interaction vertices~\cite{flex},
%$\Gamma_{\rm d}$ and $\Gamma_{\rm m}$.
The estimates are reasonable {\it vis-\`{a}-vis}
$t\!\to\!0$ limit results discussed below.
%R2

As $V_{1}$ approaches $V_{1}^{\rm md}$,
%the physical picture changes qualitatively as
CDW fluctuations rapidly take over
%R2
and due to their nearly singular nature
preclude convergence of our self-energy iteration 
for $T\!\lesssim\!0.1t$ and $V_{1}\!>\!V_{1}^{\rm md}$.
%R2
In the 1st neighbor ($n\!=\!1$) Coulomb model,
the CDW fluctuations exhibit 
maximum $\lambda^{\rm d}$
near ${\bf Q}\!=\!(\pi,\pi)$. 
They produce a repulsive 1st neighbor contribution 
to $\Gamma_{\rm s}$ {\it and}, via $\Sigma$, they
weaken the strength of the the attractive 
${\bf Q}\!\sim\!(\pi,\pi)$ spin fluctuation 
contribution to $\Gamma_{\rm s}$ when $V_{1}\!\to\!V_{1}^{\rm md}$. 
The ${\bf Q}\!\sim\!(\pi,\pi)$ charge fluctuations
also generate an on-site attraction which is too weak, however,
to overcome the on-site repulsion due to $U$ and 
${\bf Q}\!\sim\!(\pi,\pi)$ spin fluctuations.
%R2
%
%As a result,
%%all pairing eigenvalues are severely suppressed 
%no pairing instabilities of any symmetry are found 
%for $V_{1}\!>\!V_{1}^{\rm md}$,
%down to $T\!\sim\!0.1t$.
%
No pairing instabilities are found
down to $T\!\sim\!0.1t$
for $V_{1}\!>\!V_{1}^{\rm md}$;
the dominant attractive symmetries are
odd-$\omega$ $s$-wave in the triplet
and even-$\omega$ $s$-wave in the singlet
channel with respective pairing eigenvalues
$\lambda^{\rm t}\!\sim\!0.35$--$0.45$ and
$\lambda^{\rm s}\!\sim\!0.25$--$0.4$
in the $10$--$20\%$ doping range.
$d_{x^2-y^2}$ pairing is suppressed.
The singlet $s$-wave attraction
% $s$-wave pairing eigenvalue is
% attractive, with $\lambda^{\rm s}\!\sim\!0.3$, 
% suggesting 
suggests that CDW fluctuations
may enhance conventional phonon-mediated pairing.
% for $V_{1}\!>\!V_{1}^{\rm md}$.
%R2
In the 2nd neighbor ($n=2$) Coulomb model, the SDW-CDW crossover is 
similar, but driven by  CDW fluctuations
with maximum $\lambda^{\rm d}$ near ${\bf Q}\!=\!(0,\pi)$ 
and $(\pi,0)$.

The physical origin of the SDW-CDW 
crossover can be understood by considering
the exactly solvable ionic limit, $t\!\to\!0$, 
at $1\over2$-filling~\cite{ScFe}.
Large $U$ favors a groundstate with 
single occupancy at each site and AF SDW order for $|t|\!=\!0^+$.
Large $V_{1}$, in the 1st neighbor Coulomb model,
favors a CDW groundstate with modulation wavevector 
${\bf Q}^*\!=\!(\pi,\pi)$
and alternating double and zero occupancy 
in the respective two sub-lattices.
The transition between these two states occurs
when $V_{1}$ reaches~\cite{ScFe}
$V_{1}^{{\rm md}|t=0}={1\over4}U$ 
which coincides with the exact
crossover $V_{1}^{\rm md}={1\over4}U$ shown in Fig.~\ref{fig2}. 
In the 2nd neighbor Coulomb model, the competing CDW state has
${\bf Q}^*\!=\!(0,\pi)$ or $(\pi,0)$,
with alternating doubly and zero occupied rows or columns, and
the transition occurs at 
$V_{1}^{{\rm md}|t=0}\!=\!\sqrt{2}U/4\!\cong\!0.354U$.
This is roughly consistent with the approximately $T$- and 
$\langle n\rangle$-independent
crossover $V_{1}^{\rm md}\!\cong\!0.406U$ 
from Fig.~\ref{fig2}(a).
This analysis suggests that the crossover 
is driven by incipient CDW instabilities 
in the $1\over2$-filled
Mott-Hubbard insulator.
%R2
We 
%note in passing
emphasize that
such CDW instabilities 
and their effects on pairing
%R2
{\it cannot} be treated
in strong-coupling versions of the Hubbard
model, such as the $t$-$J$ model~\cite{tj_mod}. 
CDW fluctuations in the $1\over2$-filled system are 
precluded from the outset by the no-double-occupancy 
Hilbert space constraints imposed in such models.

By exploiting the retarded nature
of the pairing potential, the pair wavefunction
can evade the 
destructive effects of the instantaneous Coulomb 
repulsion~\cite{pseudo_pot}, provided
the characteristic frequency $\Omega_B$ 
of the pairing-mediating boson is much lower than
characteristic electronic energy scales such as the
Fermi energy $\epsilon_{\rm F}$.
This Coulomb ``pseudopotential'' effect~\cite{pseudo_pot}
is crucial for conventional phonon-mediated 
superconductors, where 
$\Omega_{\rm B}/\epsilon_{\rm F}\sim\!10^{-2}$--$10^{-3}$.
However, in the present problem,
$\Omega_{\rm B}/\epsilon_{\rm F}\!\sim\!10^{-1}$ 
is {\it not} very small, since the spin fluctuation
frequencies $\Omega_{\rm B}$ extend up to a
sizeable fraction of the electronic bandwidth~\cite{PaBi,sf_spect,ScNo}.
By a variational analysis,
based on restricted trial pair wavefunctions,
%\cite{ESB},
we can indeed show that the pseudopotential effect raises 
$\lambda^{\rm s}$
and $T_c$ by no more than a few percent, {\it i.e.}, it is largely
inoperative in spin fluctuation mediated pairing.

Next, we consider longer-range 
%R2
$1/r$ Coulomb terms.
% of the extended Coulomb interaction.
%R2
As shown in Fig.~\ref{fig2}(a), the inclusion of 
2nd neighbor ($n=2$) Coulomb terms suppresses
$\tc$ more strongly at low $V_{1}$ than in the 1st
neighbor model. 
%R2
However, the $d$-wave pairing also
survives to larger $V_{1}$, 
% values since $V_{1}^{\rm md}$ is
since $V_{1}^{\rm md}$ is increased in the 2nd neighbor model.
Comparable $V_{1}^{\rm md}$ 
%comparable to those in the 2nd neighbor model 
%
are found in longer-range models
with $n\!>\!2$. 
%%%%
%
% FIG. 3:
%
\begin{figure}[hbt]
\begin{picture}(3.375,2.100)
\includegraphics{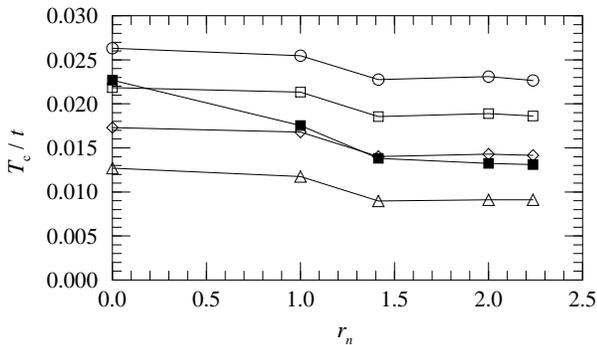}
\end{picture}
%
%%%%\begin{figure}[hbt]
%%%%\begin{picture}(3.375,2.600)
%%%%\special{psfile=fig3.eps
%%%hoffset=-43 voffset=267 vscale=50 hscale=50 angle=-90}
%%%%\end{picture}
%
%%%%
%
\caption
{
$\dxy$ $\tc$ vs.\ cutoff radius $r_n$
of the extended Coulomb interaction. Here, 
$r_{n}=0,\:1,\:\protect\sqrt{2},\:2$, and $\protect\sqrt{5}$, for
$n=0,\:1,\:2,\:3$, and $4$th neighbor cutoff, respectively.
Open symbols are for $U/t=8$, $V_{1}/t=1$;
closed symbols are for $U/t=4$, $V_{1}/t=0.5$;
with $\langle n\rangle=1.12$ (circles), $1.16$
(squares), $1.20$ (diamonds), and $1.24$ (triangles).
}
\label{fig3}
\end{figure}
%
%%%%

As shown in Fig.~\ref{fig3}, 
the dependence of $\tc$ on the cutoff radius $r_n$ at fixed $V_1$ 
rapidly saturates for $n\!\geq\!2$. 
This can be understood by noting that the long-range 
($n\!\geq\!3$) tail of the extended Coulomb potential
is strongly suppressed by the metallic screening, while the
the $d$-wave pairing wavefunction is
mostly concentrated at 1st neighbor distance. Hence the
effect of the additional ($n\!>\!2$)
direct repulsion terms on the pairing is rapidly 
reduced with increasing $r_n$.
%The basic analysis of the effects of longer-range 
%repulsions on $\Sigma$ and $\Gamma_{\rm s}$ is quite
%similar to the 1st neighbor case. 
We are thus confident that our conclusions based 
on finite-range ($n\!\leq\!4$)
models are not fundamentally altered in the
infinite-range $1/r$ model.

For comparison, we have also explored the effects of 
an added extended Coulomb repulsion on the $d$-wave $\tc$
in recently proposed phenomenological spin fluctuation
exchange models~\cite{ScNo,phen_mod}.
The $d$-wave pairing is much more rapidly suppressed with $V_{1}$
in such models, primarily due to the lack of screening,
{\it i.e.}, due to the fact that the 
spin fluctuation mediated
pairing potential, as extracted from fits 
to experimental data, is ``rigid'' and 
does not become modified by 
the extended Coulomb repulsion.

In summary, we find that, at the level of the FLEX 
approximation to the one-band Hubbard model,
spin fluctuation mediated $\dxy$ pairing
is robust against extended $1/r$
electron-electron Coulomb repulsions of realistic strengths.
The extended part of the Coulomb interaction does not cause 
a significant change in the spin fluctuation spectrum 
nor in the pairing vertex of the Hubbard model for moderate
strengths $V_{1}$.
The robustness of the model against the inclusion of extended
Coulomb terms increases with increasing on-site 
repulsion $U$.
With increasing real-space cutoff radius $r_n$, 
the extended Coulomb effects are almost entirely
saturated after the inclusion of the
2nd neighbor repulsion terms. 
A strong suppression of $\dxy$ pairing is found only when the
extended Coulomb repulsion strength $V_{1}$ becomes so large
that it induces CDW instabilities in the
$1\over2$-filled Mott-Hubbard insulator. The ubiquity
of AF SDW order in undoped insulating cuprates 
appears to rule out the latter scenario.

We thank J.~W. Serene for helpful discussions.
This work was supported in part by NSF
under grants DMR--92--15123 (H.-B.~S.) and DMR--95--20636 (N.~E.~B.).
Computing support from UCNS/RCR at UGA, from
ISD/RCF at USC,
and from NCSA at UIUC
is gratefully acknowledged.
%
%
% REFERENCES
%
%
%
%\cite{d_wave_exp}.
%\cite{d_wave_theory}.
%\cite{flex},
%\cite{HSSJ}.
%\cite{ScFe}.
%\cite{SGEH}.
%\cite{AsLa}
%\cite{EsBi}
%\cite{HaSh}
%\cite{PaBi}. 
%\cite{flex_old}.
%\cite{ov_scrn}.
%\cite{tj_mod}. 
%\cite{pseudo_pot}
%\cite{sf_spect}.
%\cite{ScNo}.
%%\cite{ESB},
%\cite{phen_mod}.
%
%

% \begin{figure}
% \caption{}
% \label{}
% \end{figure}

% \begin{table}
% \caption{}
% \label{}
% \end{tabular}
% \end{table}


\begin{references}

\bibitem[*]{email}
Email: esirgen@uga.edu

\bibitem{d_wave_exp}
For a review of the experiments see
D.~J. van Harlingen, Rev.\ Mod.\ Phys.\ {\bf 67}, 515 (1995);
and references therein.

\bibitem{d_wave_theory}
For a review of the theoretical aspects see
D.~J. Scalapino, Phys.\ Rep.\ {\bf 250}, 329 (1995);
and references therein.

\bibitem{flex} 
%For a technical introduction see
N.~E. Bickers and D.~J. Scalapino, Ann.\ Phys.\ {\bf 193}, 206 (1989).

\bibitem{HSSJ}
M.~S. Hybertsen {\it et al.}, \prb {\bf 41}, 11068 (1990);
C.-X. Chen and H.-B. Sch\"uttler, {\it ibid.}\ {\bf 43}, 3771 (1991);
S.~B. Bacci {\it et al.}, {\it ibid.}\ {\bf 44}, 7504 (1991).

\bibitem{ScFe} 
H.-B. Sch\"uttler and A.~J. Fedro,
\prb {\bf 45}, 7588 (1992).

\bibitem{SGEH}
H.-B. Sch\"uttler {\it et al.}, {\tt cond-mat/9805133}.

\bibitem{AsLa}
L.~G. Aslamazov and A.~I. Larkin, Fiz.\ Tverd.\ Tela (Leningrad)
{\bf 10}, 1104 (1968) [Sov.\ Phys.\ Solid State {\bf 10}, 875 (1968)].

\bibitem{EsBi} 
G. Esirgen and N.~E. Bickers, Phys.\ Rev.\ B {\bf 55}, 2122 (1997);
{\it ibid.}\ {\bf 57}, 5376 (1998).
%; and references therein.

\bibitem{HaSh}
W. Hanke and L.~J. Sham, \prl {\bf 33}, 582 (1974);
\prb {\bf 12}, 4501 (1975).

\bibitem{PaBi} 
C.-H. Pao and N.~E. Bickers,
Phys.\ Rev.\ Lett.\ {\bf 72}, 1870 (1994);
Phys.\ Rev.\ B {\bf 49}, 1586 (1994); 
{\it ibid.}\ {\bf 51}, 16310 (1995);
C.-H. Pao and H.-B. Sch\"uttler,
{\it ibid.}\ {\bf 57}, 5051 (1998).

\bibitem{flex_old}
N.~E. Bickers \etal,
%%, D.~J. Scalapino, and R.~T. Scalettar,
Intl.\ J. Mod.\ Phys.\ B {\bf 1}, 687 (1987); 
%
%%N.~E. Bickers, D.~J. Scalapino, and S.~R. White, 
Phys.\ Rev.\ Lett.\ {\bf 62}, 961 (1989);
Phys.\ Rev.\ B {\bf 43}, 8044 (1991).

%\bibitem{ov_scrn}
%Recent quantum Monte Carlo results suggest that the
%overscreening due to charge fluctuations in the
%doped large-$U$ 2D Hubbard system is so strong that even
%the screened on-site potential becomes attractive;
%cf. Ref.~\onlinecite{SGEH}.
%It is possible that the FLEX approximation underestimates
%this charge fluctuation screening.
%
\bibitem{tj_mod}
V.~J. Emery {\it et al.\/}, \prl {\bf 64}, 475 (1990); 
T. Barnes and M.~D. Kovarik, \prb {\bf 42}, 6159 (1990);
C. Gazza {\it et al.}, {\tt cond-mat/9803314}.

\bibitem{pseudo_pot}
D.~J. Scalapino, in {\em Superconductivity},
edited by R.~D. Parks (Marcel Dekker, New York, 1969), Vol.~1;
P. B. Allen and B. Mitrovic, in {\it Solid State Physics},
edited by H. Ehrenreich {\it et al.}\ (Academic Press, New York, 1982),
Vol.~37, p.~1--91.

\bibitem{sf_spect}
R. Preuss {\it et al.}, \prl {\bf 79}, 1122 (1997).

\bibitem{ScNo}
H.-B. Sch\"uttler and M.~R. Norman,
\prb {\bf 54}, 13295 (1996).
%
%\bibitem{ESB}
%G. Esirgen, H.-B. Sch\"uttler, and N.~E. Bickers (unpublished).

\bibitem{phen_mod}
% MMP:
A. Millis \etal,
%%, H. Monien, and D. Pines, 
Phys.\ Rev.\ B {\bf 42}, 167 (1990);
%
% BSE:
P. Monthoux and D. Pines, 
{\it ibid.}\ {\bf 49}, 4261 (1994);
%
R. J. Radtke \etal,
%%, S. Ullah, K. Levin, and M. R. Norman,
{\it ibid.}\ {\bf 46}, 11975 (1992);
%
% impurity:
%%R. J. Radtke, K. Levin, H.-B. Sch\"{u}ttler, and
%%M. R. Norman, 
%%{\it ibid.}\ {\bf 48}, 653 (1993);
%
% vHS:
%%R. J. Radtke, K. Levin, H.-B. Sch\"{u}ttler, and
%%M. R. Norman, 
{\it ibid.}\ {\bf 48}, 15957 (1993).

\end{references}
\end{document}